\documentclass[pra,twocolumn]{revtex4}
\usepackage{bbm}
\usepackage{natbib}
\usepackage{graphics}
\usepackage{amsmath}
\usepackage{mathrsfs}
\usepackage{theorem}
\usepackage{float}
\usepackage{hyperref}
\usepackage{rotating}
\usepackage{amssymb}
\usepackage[english]{babel}
\usepackage{color}
\usepackage{CJK}
\usepackage{multirow}
\usepackage{lscape}
\usepackage{fancybox}

\newcommand{\ket}[1]{|#1\rangle}

\newcommand{\inp}[2]{\langle #1 | #2\rangle}
\newcommand{\be}{\begin{eqnarray}}
\newcommand{\ee}{\end{eqnarray}}

\begin{document}
\title{Localization-enhanced dissipation in a generalized Aubry-Andr\'{e}-Harper model coupled with Ohmic baths}
\author{H. T. Cui $^{1}$}
\email{cuiht01335@aliyun.com}
\author{M. Qin $^{1}$}
\email{qinming@ldu.edu.cn}
\author{L. Tang $^{1}$}
\author{H. Z. Shen $^{2}$}
\author{X. X. Yi $^{2}$}
\email{yixx@nenu.edu.cn}
\affiliation{$^1$ School of Physics and Optoelectronic Engineering, Ludong University, Yantai 264025, China}
\affiliation{$^2$ Center for  Quantum Sciences, Northeast Normal University, Changchun 130024, China}
\date{\today}

\begin{abstract}
In this work, the exact dynamics of excitation in the generalized Aubry-Andr\'{e}-Harper model coupled with an Ohmic-type environment is discussed by evaluating the survival probability and  inverse participation ratio of the state of system. In contrast to the common belief that localization will preserve the information of the initial state in the system against dissipation into the environment, our study found that strong localization can enhance the dissipation of quantum information instead. By a thorough  examination of the dynamics, we show that the coherent transition between the energy state of system is crucial for understanding this unusual behavior. Under this circumstance, the coupling induced  energy exchange between the system and its environment can induce the periodic population of excitation on the  states of system.  As a result, the stable or localization-enhanced decaying of excitation can be observed, dependent on the energy  difference between the states of system. This point is verified in further by checking the varying of dynamics of excitation in the system when the coupling between the system and environment is more strong.
\end{abstract}

\maketitle

\section{introduction}

The closed quantum system can display resistance to the thermalization  under its own intrinsic dynamics when it is localized, e.g.,  induced by static disorder\cite{rmp}, a linear potential with a spatial gradient \cite{stark},  or the presence of a special subspace of the Hilbert space\cite{scar,fragment}. Experimental evidence for the violation of ergodicity  was presented, e.g., in ultracold atomic fermions \cite{exp-quasidisorder} and superconducting systems \cite{roushan17}. In practice, no realistic systems can be immune to environmental coupling.  Recent studies have found that even when the system is eventually  driven to the thermal equilibrium,  the localization decays  slowly \cite{opendisorder}. This strained localization  decay induces a large time window during which the nonergodic character of the system becomes apparent \cite{luschen2017}.

While localization will be detrimental to the transportation of quantum particles in systems, it was recently discovered that increasing  disorder within a system can enhance  particle transport \cite{disorderenhance}. Furthermore, combined with  environment-induced dephasing, the localized system can display robust quantum transport \cite{opendisorderenhance}. These findings imply that the interplay of  localization and environment-induced decoherence can give rise to intriguing and  complex dynamics in quantum systems. It motivates us to reconsider the robustness of localization, based on a fundamental point that the system is dissipative because of coupling to the environment. For this purpose, we studied the exact evolution of single excitation in a one-dimensional lattice system coupled to a bosonic environment. Different from Markovian treatments in previous works \cite{opendisorder}, the exact dynamics of excitation can display strong dissipation or stable oscillation,  depending on the localization of initial state. Moreover, we found that  strong localization may enhance the decaying of excitation, rather than preserve excitation in the system.  We argue that this counter institutive feature is a consequence of the energy exchange between the system and environment, which induces the coherent transition in the energy levels of system.

The work is divided into five sections. Following the preceding introduction, section II  introduces the model, and the dynamic equation for excitation is derived.  In section III, we  discusses the time evolution of excitation for different cases by calculating the survival probability of the information of initial state and the inverse participation ratio, both of which characterize the  localization of system. Subsequently,  a physical explanation is provided  in section IV. It is shown that the energy exchange between the system and environment is responsible for the unusual observation.  Finally, conclusion is presented in section V.

\section{The model and dynamic equation }

In this work, we focus on the open dynamics of single excitation in a generalized Aubry-Andr\'{e}-Harper (GAAH) model described by the  Hamiltonian below \cite{gAAH}
\be\label{hs}
H_S=\lambda \sum_{n=1}^N \left(c^{\dagger}_{n} c_{n+1} +c^{\dagger}_{n+1} c_{n} \right) + \nonumber \\
\Delta \frac{\cos(2\pi \beta n +\phi)}{1 - a \cos(2\pi \beta n +\phi)}c^{\dagger}_n c_{n},
\ee
where $N$ denotes the number of lattice sites and $c_n (c^{\dagger}_n)$ is the annihilation (creation) operator of excitation at the $n$-th lattice site. For a quasi-periodic modulation, we adopt $\beta=\left(\sqrt{5}-1\right)/2$ with respect to the recent experimental verification of delocalization- localization transition \cite{exp-quasidisorder}. The onsite potential is a smooth function of parameter $a$ in open interval $a\in(-1, 1)$. When $a=0$, Eq. \eqref{hs} reduces to the standard  AAH model \cite{aah}, in which a delocalization- localization phase transition can  occur when $\Delta=2\lambda$. Whereas for $a\neq 0$, GHHA exhibits an exact mobility edge (ME) following the expression \cite{gAAH}
\be\label{me}
a E_c = \text{sign}\left(\lambda\right)\left(2\left|\lambda\right| - \left|\Delta\right|\right).
\ee
In the above, $E_c$ is a special eigenenergy of $H_S$, that separates the extended eigenstates from localized counterparts. The coexistence of localized and delocalized states is a typical feature of GAAH model, and leads to the complex  excitation  dynamics in the system. To avoid the boundary effect, the periodic boundary condition, i.e., $c_n=c_{n+N}$,  is adopted. Since the current work focuses on the robustness of localization in the system, $\phi=\pi$ is adopt without a loss of generality.  For simplicity, $\hbar=\lambda \equiv 1$ is assumed  in the following discussion. Recently, the localization properties of GAAH model has been experimentally investigated in optical lattices \cite{alexan21}.

To establish the open dynamics of localization in the GAAH model, bosonic reservoir with different modes characterized by frequencie $\omega_k$ are introduced as the environment. Its Hamiltonian can be written as follows:
\be
H_B= \sum_k \omega_k b_k^{\dagger}b_k,
\ee
where $b_k$ ($b_k^{\dagger}$) is the annihilation (creation) operator of reservoir $k$.  The system is coupled to the environment via particle-particle exchanging,
\be
H_{int}= \sum_{k, n}\left(g_k b_k c_n^{+} + g_k^* b_k^{\dagger} c_n\right)
\ee
where $g_k$ is the coupling amplitude between the system and reservoir mode $k$. The complexity of dynamics is determined by the spectral density
\be
J(\omega)= \sum_k \left| g_k \right|^2 \delta\left(\omega- \omega_k\right),
\ee
which characterizes the energy structure of the system plus the system-environment interaction. In this work, Ohmic-type spectral density is adopt as follows:
\be\label{j}
J(\omega)= \eta \omega \left(\frac{\omega}{\omega_c}\right)^{s-1}e^{-\omega/\omega_c}.
\ee
The quantity $\eta$ is simply the classical friction coefficient, and thus forms a dimensionless  measure of the coupling strength between the system and its environment \cite{leggett}. Actually, Eq. \eqref{j} characterizes the damping movement of electrons in a potential, and can simulate a large class of environments. The environment can be classified  as sub-Ohmic ($s<1$), Ohmic ($s=1$) or super-Ohmic ($s>1$) \cite{leggett}. Without loss of generality, we focus on the Ohmic case ($s=1$) since it characterizes the typical dynamics of dissipation in the system \cite{leggett}. Here, $\omega_c$ is the cutoff frequency of the environment spectrum, beyond  which the spectral density starts to fall off; hence, it determines the regime of reservoir frequency, which is dominant for dissipation. In general, the value of $\omega_c$ depends on the specific environment. In this work, $\omega_c=10$ is set in order to guarantee that the highest energy level in $H_S$ is embedded into the continuum of the environment.

With respect of the absence of particle interaction in Eq.\eqref{hs}, it is convenient to focus our attention on the dynamics of single excitation. Under this prerequisite, the environment is set to be at zero temperature such that it can be at vacuum state.  With respect of the environment at finite-temperature, the recent studies have shown that the localization may be destroyed  by heating dynamics. As a result, the system relaxes eventually into a infinite-temperature state \cite{opendisorder}. However, the relaxation is  logarithmically  slow in time because of the localization in system \cite{opendisorder}. This means that there is a  time window where the nonergodic feature of localization can be observed in experiments \cite{luschen2017} . So  it is expected that the dynamics of localization can display some unusual feature at zero temperature because of the strong coupling between system and environment. Furthermore, the dynamics of localization can be treated exactly in this case. And thus one can get a comprehensive picture how the localization in the system is varied by coupling to environment.  For this purpose,  the dynamical  equations of excitation must be derived first. Formally, at any time $t$,  the state of the  system plus environment can be written as
\be\label{state}
\ket{\psi(t)}&=& \left(\sum_{n=1}^N \alpha_n(t) \ket{1}_n\ket{0}^{\otimes (N-1)} \right)\otimes \ket{0}^{\otimes M} + \nonumber \\ &&\ket{0}^{\otimes N} \otimes \left(\sum_{k=1}^{M} \beta_k(t) \ket{1}_k\ket{0}^{\otimes (M-1)} \right),
\ee
where $\ket{1}_n = c_n^{\dagger}\ket{0}_n$ denotes the occupation of the $n$-th lattice site, $\ket{0}_k$ is the vacuum state of $b_k$ and $\ket{1}_k=b_k^{\dagger}\ket{0}_k$, and $M$ denotes the number of  mode in the environment. Substituting Eq. \eqref{state} into Schr\"{o}dinger equation and solving first for $\beta_k (t)$, one can get a  integrodifferential equation for $\alpha_n(t)$,
\be\label{evolution}
\mathbbm{i}\frac{\partial }{\partial t}\alpha_n(t)&=& \left[\alpha_{n+1}(t) + \alpha_{n-1}(t)\right]+ \Delta \cos(2\pi \beta n +\phi)\alpha_n(t) \nonumber\\
&&- \mathbbm{i} \sum_{n=1}^N \int_0^t \text{d}\tau\alpha_n(\tau)f(t-\tau),
\ee
where $\mathbbm{i}$ is the square root of $-1$, and the  memory kernel $f(t-\tau)$ is defined as
\be
f(t-\tau)= \int_0^{\infty} \text{d} \omega J(\omega)e^{-\mathbbm{i} \omega(t-\tau)}.
\ee
It is evident that the population amplitude  of $\alpha_n(t)$ is  significantly correlated to its past values. For  spectral density shown in Eq. \eqref{j}, $f(t-\tau)= \frac{\eta}{\omega_c^{s-1}} \frac{\Gamma(s+1)}{\left[\mathbbm{i}(t-\tau) + 1/\omega_c\right]^{s+1}}$.  It is noted that the Markovian limit could be obtained by replacing $\alpha_n(\tau)$ in Eq. \eqref{evolution} with its current value $\alpha_n(t)$. As a result, the last term on the right hand of Eq. \eqref{state} contributes a positive term to Eq. \eqref{evolution}, which depicts the decaying of $\alpha_n(t)$ \cite{hnxiong10}.  So, it is expected that the non-Markovian feature would  inspire the distinct dynamics of excitation.

\section{Dynamical evolution of excitation in the system}
\begin{figure*}
\center
\includegraphics[width=12cm]{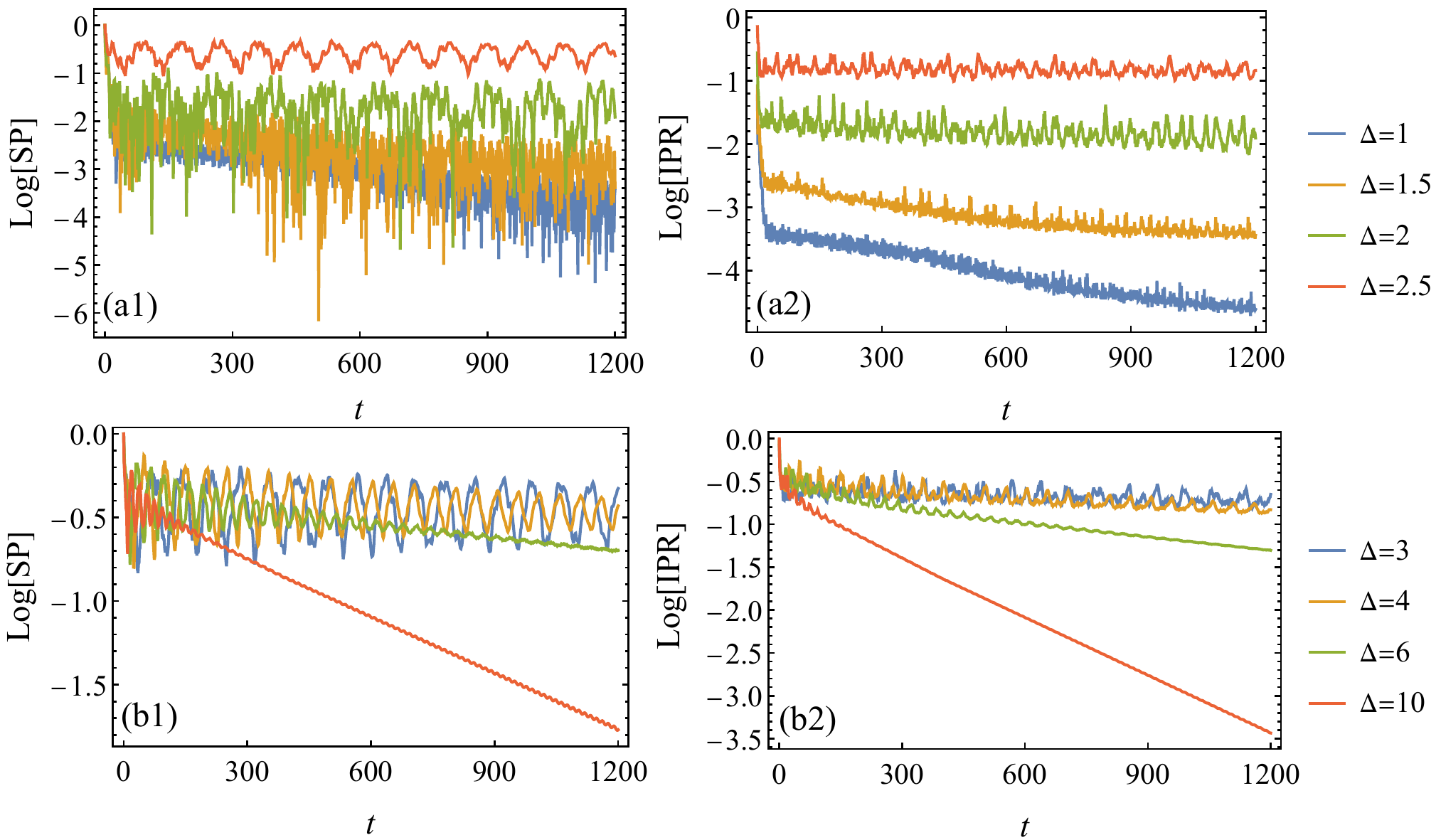}
\caption{(Color online) Logarithmic plotting for the time evolution of SP (right column) and IPR (left column) when $a=0$ for different values of  $\Delta$. The initial state is chosen as the highest excited state of $H_S$ in all plots. For these plots, $\eta=0.1$, $\omega_c=10$, $s=1$ and $\phi=\pi$ are chosen. }
\label{fig:a=0}
\end{figure*}

Taking the existence of ME into account, we focus on the evolution of excitation initially in the  highest excited eigenstate (ES) of $H_S$ \cite{alexan21}. Thus, as for Eq. \eqref{me}, the higher eigenenergy refers to the stronger localization in the ES \cite{gAAH}. Moreover,  since ES is embedded in the continuum of the environment, the occurrence of a bound state is excluded from the current discussion \cite{john}. We introduce the survival probability (SP), defined as $SP=\left|\inp{\psi(t)}{ES}\right|^2$,  to characterize the  dissipation of quantum information. In addition, the inverse participation ratio (IPR), defined as $\text{IPR}=\sum_{n=1}^N \left|\alpha_n(t)\right|^4$, is also calculated to establish localization variation in the system. Both SP and IPR have been  extensively used  to explore the dynamics of localization in disordered many-body systems \cite{hs15}.

The following discussion focuses on two cases, i.e., $a=0$ and $a=0.5$. For the former case, ME is absent. However, a delocalization- localization transition can occur in the system when $\Delta=2$, which separates the delocalized ($\Delta<2$) from the localized  ($\Delta>2$) phase. This transition is absent in the latter; instead,  the eigenstates are classified as localized  ($E>E_c$) or extended ($E<E_c$) since the occurrence of ME.

Finally, the difficulty of finding the analytical solution to  Eq. \eqref{evolution} is noted. Thus, we have to rely on numeric method. Our method is to transform the integral in  Eq. \eqref{evolution}  into a summation with properly chosen step length. By  solving Eq. \eqref{evolution} iteratively, $\alpha_n(t)$ could be determined for any times $t$. However, the computational cost grows exponentially as the number of steps and  lattices number $N$.  To disclose the long-time behavior of SP and IPR,  $N$ is restricted to 21 so that $t=1,200$ could be achieved in a moderate  computational cost.

\subsection{$a=0$}

In Fig. \ref{fig:a=0}, the time evolution of SP and IPR is plotted for different values of $\Delta$. As shown in Fig.\ref{fig:a=0}(a1) and (a2),  both SR and IPR indicate a rapid decaying when the system is in a delocalized phase ($\Delta<2$). With an increase of $\Delta$, the decay of IPR  becomes very slow, as shown in Fig.\ref{fig:a=0}(a2). Meanwhile, a stable oscillation is developed for SP, as indicated for $\Delta=2.5$ in Fig. \ref{fig:a=0}(a1) and $\Delta=3$ and $4$ in Fig. \ref{fig:a=0}(b1). Whereas ES is localized in these cases, it manifests the robustness of localization against dissipation. This finding is different from the observation in Refs.\cite{opendisorder}, where the localized system eventually  reflect a thermal equilibrium because of coupling to the environment. It can be attributed to the effect of memory kernel $f(t-\tau)$, which makes the past and current state of system  interfered.  The stability of SP or IPR can also be manifested by finding the variance of the position of excitation within the atomic site,  $\langle\delta^2 n\rangle$. As show in Fig.\ref{fig:variance}(b) in Appendix, $\langle\delta^2 n\rangle$ displays a regular oscillation for $\Delta=2.5$. In contrast, it becomes irregular for $\Delta=1$ shown in Fig.\ref{fig:variance}(a).

With a further increase in  $\Delta$, we find  that both SP and IPR reflect significant decaying, as shown for $\Delta=6$ and $10$ in Fig.\ref{fig:a=0}(b1) and (b2). This unusual feature indicates that the strong localization may have enhanced the loss of quantum information, rather than  preserving  it against decoherence. However, we also note that, in contrast to the super-exponential decaying of the extended state, the strongly localized state decays exponentially instead. Experimentally, this implies  that it is still possible to differentiate the extended  from the localized phase by checking the process of decoherence. In addition, it is noted that $\langle\delta^2 n\rangle$ can tend to be steady shown in Fig.\ref{fig:variance}(c). This observation can be considered as a result of the strong localization  in the system.

\subsection{$a=0.5$}

\begin{figure*}
\center
\includegraphics[width=12cm]{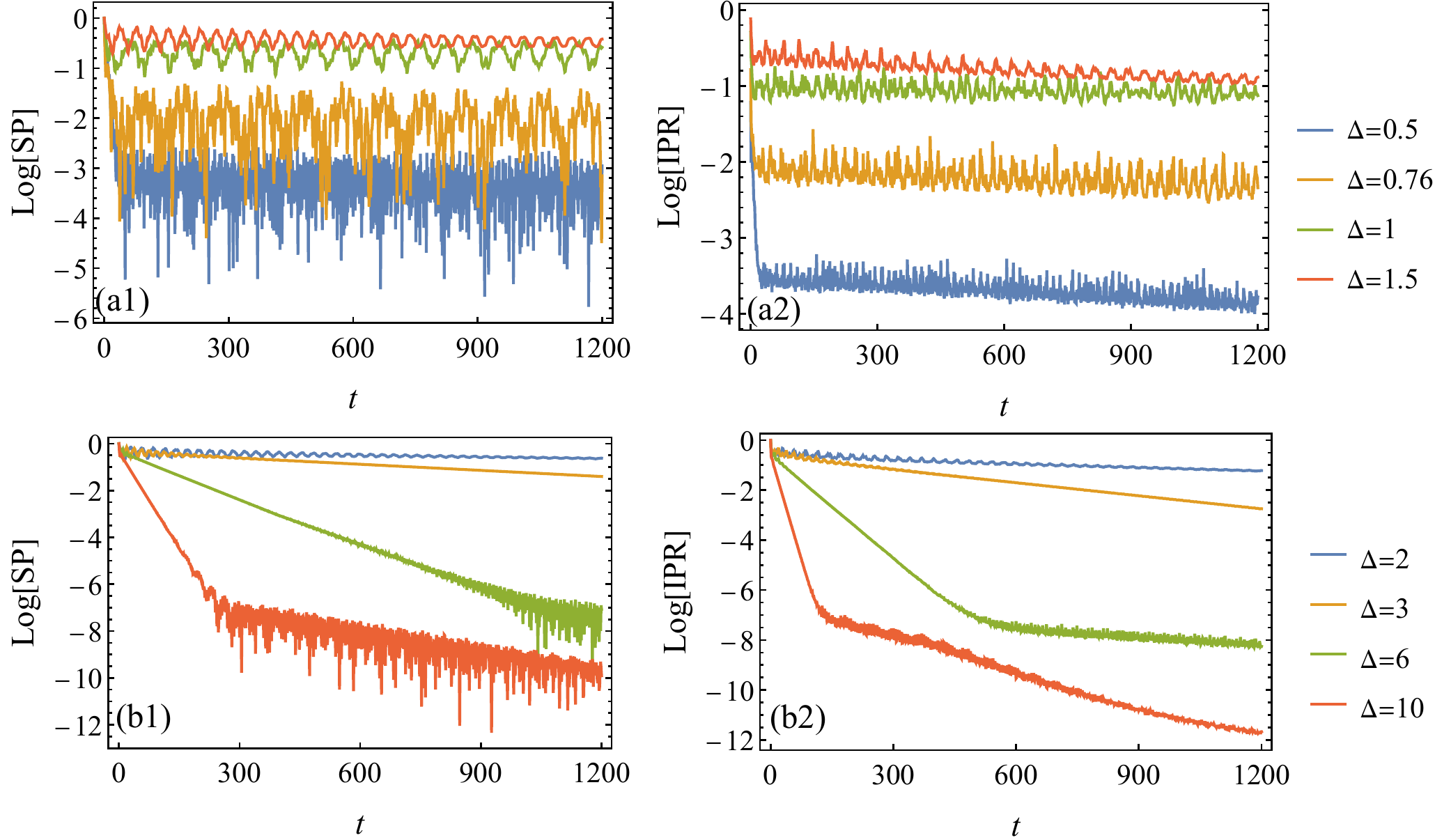}
\caption{(Color online) Logarithmic plotting for the time evolution of SP (right column) and IPR (left column) when $a=0.5$ for different values of  $\Delta$. The initial state is chosen as the highest excited state of $H_S$ in all plots. The other parameters are the same as those shown in Fig. \ref{fig:a=0}. }
\label{fig:a=0.5}
\end{figure*}

To gain a further understanding of the localization-enhanced dissipation, this section focuses on the case of $a\neq 0$. A distinct  feature in this situation is the occurrence of ME \cite{gAAH}. Consequently, the energy ES of $H_S$ may behave in a localized or extended manner, which will be decided by the relationship of  eigenvalue $E$ and $E_c$ in Eq. \eqref{me}. As a result, the system  cannot  simply be  classified as either localized or extended.  It is noted that, depending on the sign of $a$,  the maximally localized state can be exchanged between the  ground state  and the highest excited state of $H_S$  \cite{gAAH}. Since this work focuses on the interplay of localization and dissipation, $a=0.5$ is selected as an exemplification. Accordingly, ES has the largest localization. For $a<0$, the ground state is maximally localized instead. For $\omega_k>0$,  the coupling to the environment renormalizes the ground state as a dissipationless bound state \cite{john}. This unique situation is excluded from our discussion.

In Fig. \ref{fig:a=0.5}, the evolution of SR and IPR are presented for different values of $\Delta$. Two selected cases, i.e.,  $\Delta=0.5$ and $0.76$  are first studied, for which ES is extended or has the eigenvalue very closed to $E_c$. Both SP and IPR display rapidly decaying at an earlier time before very slowly descending, as shown in Fig. \ref{fig:a=0.5} (a1) and (a2). With the increase of $\Delta$, ES behaves in a more localized manner. As is anticipated, it is noted that both  SP and IPR decay slowly when $\Delta=1.5, 2$ and $3$. In contrast, a steep descent is observed when $\Delta=3, 6$ and $10$, as shown in Fig. \ref{fig:a=0.5} (b1) and (b2). This feature  is consistent with the observation in the case of $a=0$. The strong localization of a state can enhance the lost of information within the  initial state.

Similar to the former case, one can find that  both SP and IPR seem to become numerically stable when $\Delta=1$,  as shown in  Fig. \ref{fig:a=0.5} (a1) and (a2). Meanwhile, a regular oscillation  can also be noted for SP.  The stability can also be demonstrated by $\langle\delta^2 n\rangle$. As shown in Fig.\ref{fig:variance}(e) in Appendix, a regular oscillation of $\langle\delta^2 n\rangle$ can be observed. In contrast, this picture is absent for the other value of $\Delta$, as exemplifications in Fig. \ref{fig:variance}(d) and (f).

To conclude, we have shown that the localization does not always preserve the excitation in the system. Instead, strong localization enhances the decaying of excitation into the environment. However, one can also note that before this picture occurs, a stability of SP or IPR can be found. This observation implies that the localization enhanced dissipation could have different underlying physical reason  that will be disclosed in the next section.

\section{Coupling induced coherent transition among energy level $E$s}

\begin{figure*}
\center
\includegraphics[width=17cm]{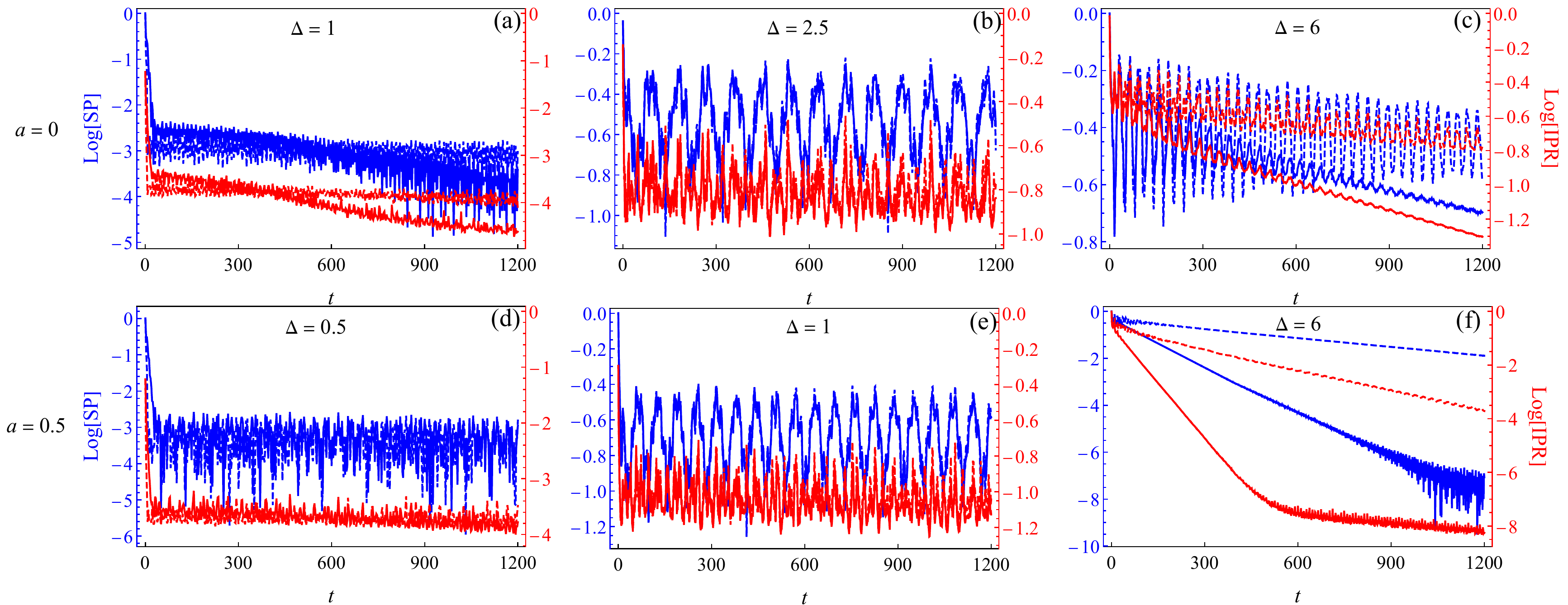}
\caption{(Color online) Comparative plotting of the time evolution of SP (blue color) and IPR (red color) for $\eta=0.1$ (solid line) and $\eta=0.5$ (dashed line). Except for $a$ and $\Delta$, $\omega_c=10$, $s=1$ and $\phi=\pi$ are selected for all plots.  }
\label{fig:eta}
\end{figure*}

In general, the periodic oscillation of SP is a manifestation of the coherent transition between two orthogonal states. To verify this point, it is necessary to find the eigenstates by solving the  Schr\"{o}dinger equation as follows:
\be\label{h}
H\ket{\psi_E}= E \ket{\psi_E},
\ee
in which $H=H_S + H_B +H_{int}$ is the total Hamiltonian of the system plus its environment. Formally, the eigenstate can be written as  $\ket{\psi_E}= \left(\sum_{n=1}^N \alpha_n \ket{1}_n\ket{0}^{\otimes (N-1)} \right)\otimes \ket{0}^{\otimes M} +\ket{0}^{\otimes N} \otimes \left(\sum_{k=1}^{M} \beta_k \ket{1}_k\ket{0}^{\otimes (M-1)} \right)$. Substituting $\ket{\psi_E}$ into Eq. \eqref{h} and eliminating $\beta_k$, we get
\be\label{psie}
\left(\alpha_{n+1} + \alpha_{n-1}\right) &+& \Delta \cos(2\pi \beta n +\phi)\alpha_n  +\nonumber \\
&& \int_0^{\infty} \text{d}\omega\frac{J(\omega)}{E-\omega} \sum_{n=1}^N \alpha_n = E \alpha_n.
\ee
By solving Eq.\eqref{psie}, the eigenenergy $E$ and the corresponding coefficient $\alpha_n$ can be determined, which characterizes the state of system after tracing out the environment. In this sense, we call the state $\sum_n \alpha_n \ket{1}_n$ as the reduced energy eigenstate of system. As shown in Appendix, because of the integral  $ \int_0^{\infty} \text{d}\omega\frac{J(\omega)}{E-\omega}$,  $E$ can be complex, and thus Eq.\eqref{psie} depicts the nonunitary dynamics of single excitation in the atomic chain. Formally, one can introduce an effective non-Hermitian Hamiltonian to characterize the nonunitary dynamics. In this point, the state $\sum_n \alpha \ket{1}_n$ can be considered as the eigenstate of the non-Hermitian Hamiltonian. However, it is difficult to construct the counterpart in this case  since $E$ is involved in  Eq. \eqref{psie}. Thus, in order to find $E$ and corresponding $\alpha_n$s, one has to resort to numerical method. 

In Appendix, Eq. \eqref{psie} is solved numerically for $a=0, \Delta=2.5$ and $a=0.5, \Delta=1$ respectively. Focusing on two highest $E$,  we found that  the difference of their real parts is consistent with the oscillation frequency  of SR, observed in previous section. For instance, it is found for $a=0$ and $\Delta=2.5$ that the oscillation of SR has a frequency of  $\sim 2\pi/77.8=0.08076$.  Correspondingly, the calculated difference between the real parts of two highest $E$s is  $2.952238 - 2.882305=0.069933$, as shown in Fig. \ref{fig:eta}(a). Similar observation can also be found  for $a=0.5$ and $\Delta=1$, that SR shows an oscillation with frequency $\sim 2\pi/57.6=0.109083$. The calculated difference is $2.705113- 2.605926=0.099187$, as shown in Fig. \ref{fig:eta}(b). The small discrepancy can be attributed to the  error of estimation.  Moreover, it is noted that  the imaginary part of the highest $E$ has a  order of  magnitude  $\sim 10^{-6}$ in both cases. This means that the decaying of SP and IPR is  slow  and neither display a discernable descent in numerical simulation up to $t\sim 10^3$. In further, we also calculate the overlap between the initial state of system and the state $\ket{\psi_{ES}}$ corresponding to the highest $E$. For $a=0, \Delta=2.5$, $\left|\inp{ES}{\psi_{ES}}\right|^2=0.498776$, which is consistent to the extremal  maximum of SP $0.496809$ extracted from the data in Fig. \ref{fig:a=0}. While for $a=0.5, \Delta=1$,  $\left|\inp{ES}{\psi_{ES}}\right|^2=0.4471716$, which is consistent to the extremal maximum of SP 0.405453  extracted from the data in Fig. \ref{fig:a=0.5}.

We have shown that the steady oscillation of SP comes from the coherent transition of system between two highest excited states. Physically, the transition is a consequence of the energy exchange between the system and its environment. Thus it is expected that the transition can be varied   by changing the coupling strength $\eta$.  A comparative study of SP or IPR  is presented  for $\eta=0.1$ (the solid line) and $\eta=0.5$ (the dashed line) in Fig. \ref{fig:eta}. It is clear  that SP or IPR may show distinct response to the increasing of  $\eta$, depending on the localization of initial state. For initial state being extended or weakly localized, the evolution of both  SP and IPR display small variation for different $\eta$, as shown in Fig.\ref{fig:eta} (a) and (d). In contrast, for initial state being highly localized,  a significant variation of SP and IPR can be observed. As shown in Figs.\ref{fig:eta} (c) and (f), the evolution of SP or IPR can be stretched significantly by increase of $\eta$. Especially, an oscillation is developed with increase of $\eta$, as demonstrated in Fig.\ref{fig:eta} (c). However, for the steady cases displayed in the previous section, the increase of coupling strength has no noticeable effect on SP or IPR, as shown in Figs.\ref{fig:eta} (b) and {e}. In order to explain this phenomenon, we also examine the two highest $E$s for different $\Delta$ or $\eta$, as shown in Table \ref{table:eta}. It is noted that the difference between real parts of the two highest $E$s is enlarged with increase of $\Delta$ or $\eta$.

Combined with these observations together, one can conclude that the localization enhanced decaying of excitation is a consequence that the environment is unable to  provide suitable energy  to drive the coherent transition of excitation between the states corresponding to the two highest $E$,  since their energy gap  become large with increase of $\Delta$. As a result,  the energy would unidirectionally flow into the environment, and the excitation may be absorbed completely by environment. Whereas for $ES$ being extended or weakly localized, the energy gap is small. Thus, the environment provides excessive energy such that make $ES$  interfered  with  more than one energy level of system. As a result, the excitation might be populated in the entire atomic sites, and the information of initial state becomes erased rapidly.  The increase of $\eta$ enhances the energy exchange between the system and its environment. This is the reason that SP or IPR becomes stretched as shown in Figs.\ref{fig:eta} (c) and (f).  

\begin{table}
\begin{tabular}{c|c|c}
   \multicolumn{3}{c}{$a=0$}\\
  \hline  \hline
  $\eta$ & 0.1 & 0.5 \\
  \hline
  \multirow{2}*{$\Delta=1$}  &  $2.0671 - 1.4187\times 10^{-10} \mathbbm{i}$  & $2.0671 - 2.8230\times 10^{-11} \mathbbm{i}$  \\  &  $2.0591 - 5.8097 \times 10^{-8}\mathbbm{i}$  & $2.05910 - 1.1555\times 10^{-8} \mathbbm{i}$\\
  \hline
  \multirow{2}*{$\Delta=2.5$} & $2.9522 - 5.0623\times 10^{-6} \mathbbm{i}$  &  $2.9522 - 1.0123 \times 10^{-6} \mathbbm{i}$   \\  & $2.8823 - 5.3124\times 10^{-5} \mathbbm{i}$ & $2.8822 - 1.0584\times 10^{-5} \mathbbm{i}$  \\
   \hline
  \multirow{2}*{$\Delta=6$} &$6.1522 - 3.3326\times 10^{-4} \mathbbm{i}$   &  $6.1519 - 6.7207 \times 10^{-5} \mathbbm{i}$  \\  & $5.9587 - 4.0397 \times 10^{-3} \mathbbm{i}$   &  $5.9539 - 8.2535\times 10^{-4} \mathbbm{i}$ \\
  \hline\hline
\end{tabular}

\begin{tabular}{c|c|c}
   \multicolumn{3}{c}{$a=0.5$}\\
  \hline  \hline
  $\eta$ & 0.1 & 0.5 \\
  \hline
  \multirow{2}*{$\Delta=0.5$}  &  $ 2.1666 - 2.7500\times 10^{-8} \mathbbm{i}$  & $2.1666 - 5.4767\times 10^{-9} \mathbbm{i}$  \\  &  $ 2.1365 - 5.3164\times 10^{-8}\mathbbm{i}$  & $  2.1365 - 1.0566\times 10^{-8} \mathbbm{i}$\\
  \hline
  \multirow{2}*{$\Delta=1$} & $ 2.7051- 7.0911\times 10^{-6} \mathbbm{i}$  &  $ 2.7051 - 1.4174\times 10^{-6} \mathbbm{i}$   \\  & $ 2.6059 - 6.0298\times 10^{-5} \mathbbm{i}$ & $ 2.6058 - 1.1989\times 10^{-5} \mathbbm{i}$  \\
   \hline
  \multirow{2}*{$\Delta=6$} &$ 11.9802 - 0.01578\mathbbm{i}$   &  $  11.9766 - 3.2128\times 10^{-3} \mathbbm{i}$  \\  & $ 11.3612- 0.1374 \mathbbm{i}$   &  $ 11.3003 - 0.0305 \mathbbm{i}$ \\
  \hline\hline
\end{tabular}
\caption{\label{table:eta} A comparison of the two highest $E$ when $\eta=0.1$ and $\eta=0.5$. The value of $E$ is determined numerically by solving Eq. \eqref{psie}. The other parameters are same as those in Fig. \ref{fig:eta}}
\end{table}

\section{Conclusion}

In conclusion, the exact dynamics of excitation, initially embedded in the highest excited state of a GAAH model coupled to an Ohmic-type environment, is studied by evaluating SP and IPR.  An important observation is that  the stable oscillation and the enhanced decaying of SP and IPR can be detected  when the localization of the initial state is moderate or strong.  This finding is distinct from the common perception that the localization in the system would protect quantum information against dissipation. To gain a further understanding of this result, the eigen energy $E$ is determined analytically by solving Eq, \eqref{psie}. It is shown that  the stable oscillation of SP and IPR  is a result of  the  coherent transition of excitation between the states corresponding the two highest $E$s, which  stems from the energy exchange between the system and the environment. Consequently,  the environment can feed back appropriate energy into the system, which induces a periodic population of the system on the two states.  However,  with the substantial increase of $\Delta$, the  energy difference between these two states is enlarged. Thus, the environment could not feed back enough energy to give rise to population. In contrast, when the energy difference is small for, the feedback of energy from environment would make the highest level interfered with the other levels. As a consequence, the information of initial state is eventually erased.  This explanation is further verified  by establishing the influence of the coupling strength $\eta$. As shown in Fig. \ref{fig:eta} (c) and (f), the increase of $\eta$   significantly  stretches  decaying  of both SP and IPR when the highest energy level $E$ is strongly localized. Whereas for the highest energy level being extended or weakly localized, the increase of $\eta$ generally enhances the decaying of  SP and IPR at a short time, as shown in Fig. \ref{fig:eta} (a) and (d).

Finally, it should be pointed out that the appearance of particle interaction can significantly modify the dynamics of excitation in the system. However, the exact treatment of open dynamics in the context of interacting many-body systems is a challenging task. Recent studies on disordered many-body systems showed that the role of interaction is to delocalize the system \cite{rmp, exp-quasidisorder}. In this sense, the interaction could act as an environment, which thermalizes the system.  Concerning the fact  that the localization-enhanced dissipation is state-dependent only, it may still be observed even if the interaction happens.

\section*{Acknowledgments}
HTC acknowledges the support of Natural Science Foundation of Shandong Province under Grant No. ZR2021MA036. MQ acknowledges the support of National Natural Science Foundation of China (NSFC) under Grant No. 11805092 and  Natural Science Foundation of Shandong Province under Grant No. ZR2018PA012. HZS acknowledges the support of NSFC under Grant No.11705025. XXY acknowledges the support of NSFC under Grant No. 11775048.

\renewcommand\thefigure{A\arabic{figure}}
\renewcommand\theequation{A\arabic{equation}}
\setcounter{equation}{0}
\setcounter{figure}{0}

\section*{Appendix}

The integral $ \int_0^{\infty} \text{d}\omega\frac{J(\omega)}{E-\omega}$  in Eq. \eqref{psie} is  divergent  when $E>0$. Therefore, to determine $E$, we apply the Sokhotski-Plemelj (SP) formula to evaluate the above integral. The SP formula can be given as
\be\label{sp}
\lim_{\epsilon\rightarrow 0} \frac{1}{x-x_0+\pm \mathbbm{i} \epsilon} = \text{P} \frac{1}{x- x_0} \mp \mathbbm{i}  \pi \delta\left(x - x_0\right),
\ee
in which P  denotes the principle value of Cauchy. Thus,
\be
\lim_{\epsilon\rightarrow 0}  \int_0^{\infty} \text{d}\omega\frac{J(\omega)}{\omega- E - \mathbbm{i} \epsilon} = \text{P}   \int_0^{\infty} \text{d}\omega\frac{J(\omega)}{\omega- E } +  \mathbbm{i} \frac{\pi}{2} J(E).
\ee
In the above derivation, the case of $- \mathbbm{i} \epsilon$ is selected. As shown in the following discussion, this choice guarantee  $E$ has a negative imaginary part that characterizes dynamic dissipation.

\begin{figure*}
\center
\includegraphics[width=8.5cm]{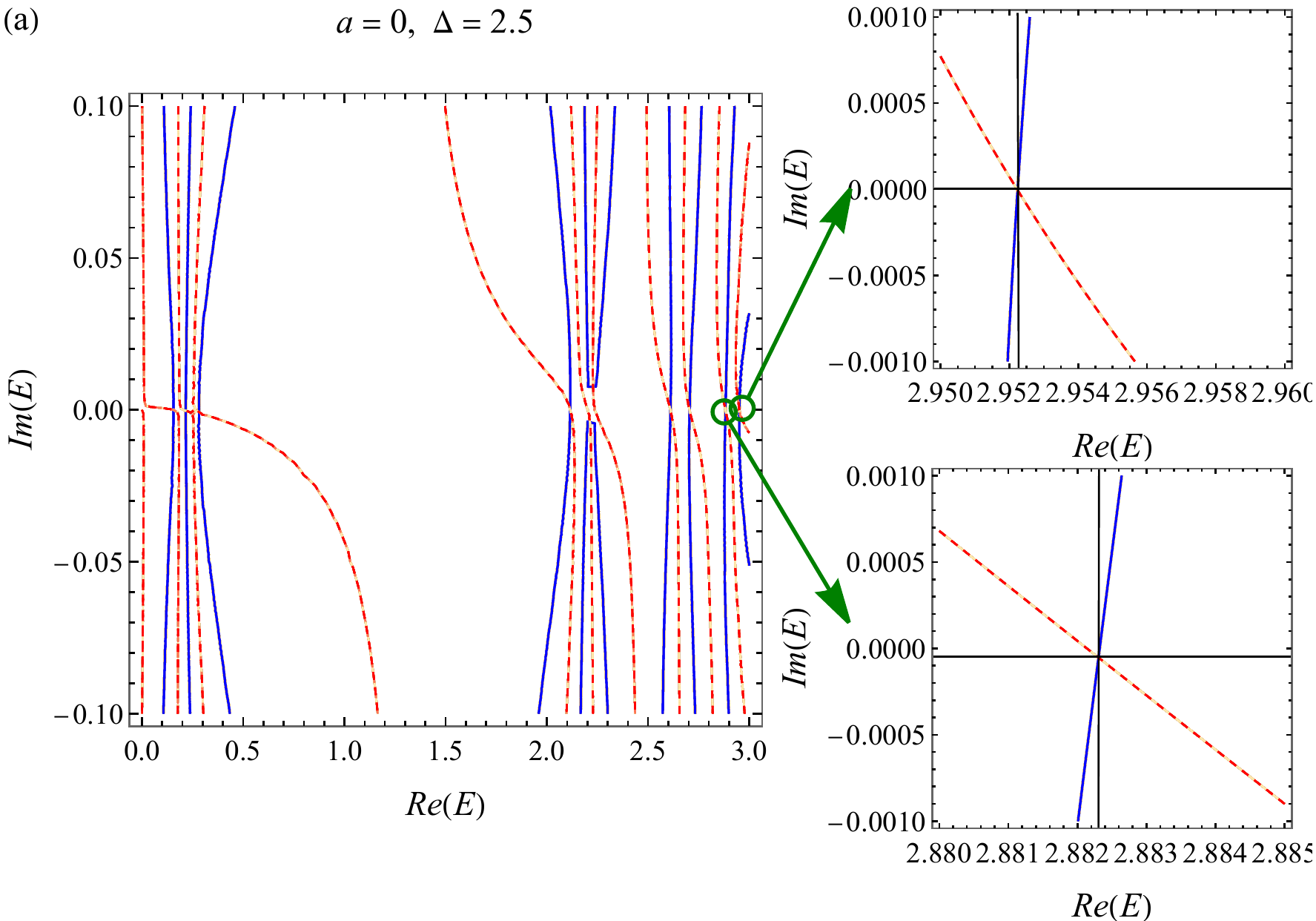}
\includegraphics[width=8.5cm]{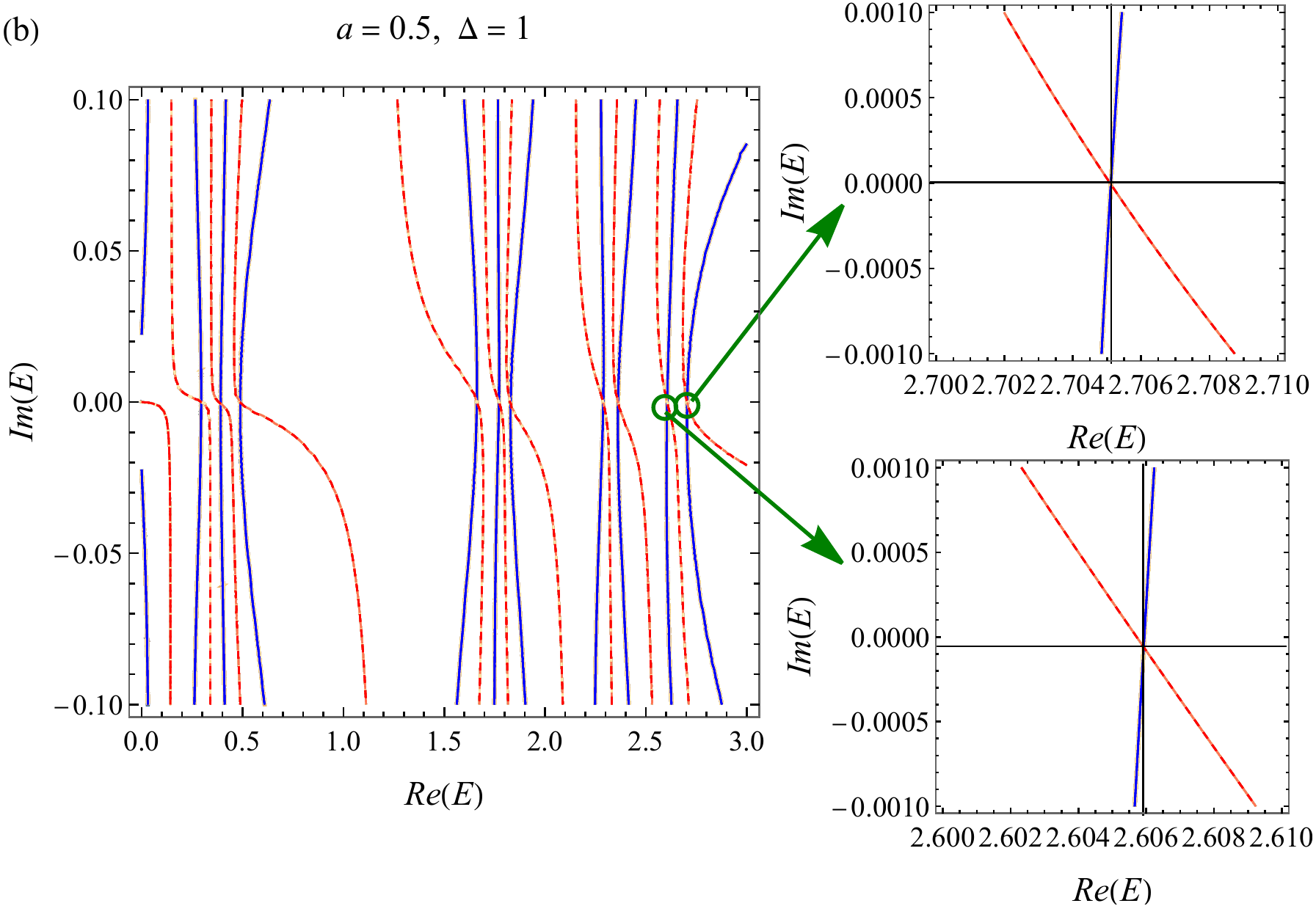}
\caption{(Color online) The  contour plots of numerical determination of $E$ in Eq. \eqref{psie} when (a) $a=0, \Delta=2.5$  and (b) $a=0.5, \Delta=1$. The solid-blue and dashed-red line panels represent respectively the vanishing real and imaginary part of the coefficient determinant in Eq. \eqref{psie}. The inner panels in (a) and (b) decide the two highest reduced energy levels.  $\eta=0.1$, $\omega_c=10$, $s=1$ and $\phi=\pi$ are chosen for these plots }
\label{fig:psie}
\end{figure*}

The value of $E$  can be established by finding zero coefficient determinants in  Eq. \eqref{psie}.  In Fig. \ref{fig:psie}, the contour plots for the vanishing real (solid- blue line) and imaginary (dashed- red line) part of coefficient determinant are presented, respectively, for (a) $a=0, \Delta=2.5$ and (b) $a=0.5, \Delta=1$, in which the crossing point of the two lines corresponds to the value of reduced energy $E$. The two inner panels in each plot show the details for the two $E$s  with the largest real part. The rigorous  value of $E$ and the corresponding  $\alpha_n$ can  be decided by recurrently solving the eigenvalue equation Eq. \eqref{h}. By doing so, the result shows that  $E_1= 2.952238 - 5.062298\times 10^{-6} \mathbbm{i}$, $E_2= 2.882305 -5.312399\times 10^{-5} \mathbbm{i}$ for case (a) and  $E_1= 2.705113 - 7.091364\times 10^{-6} \mathbbm{i}$, $E_2=2.605926 - 6.029\times 10^{-5} \mathbbm{i}$ for case (b).

\begin{figure*}
\center
\includegraphics[width=15cm]{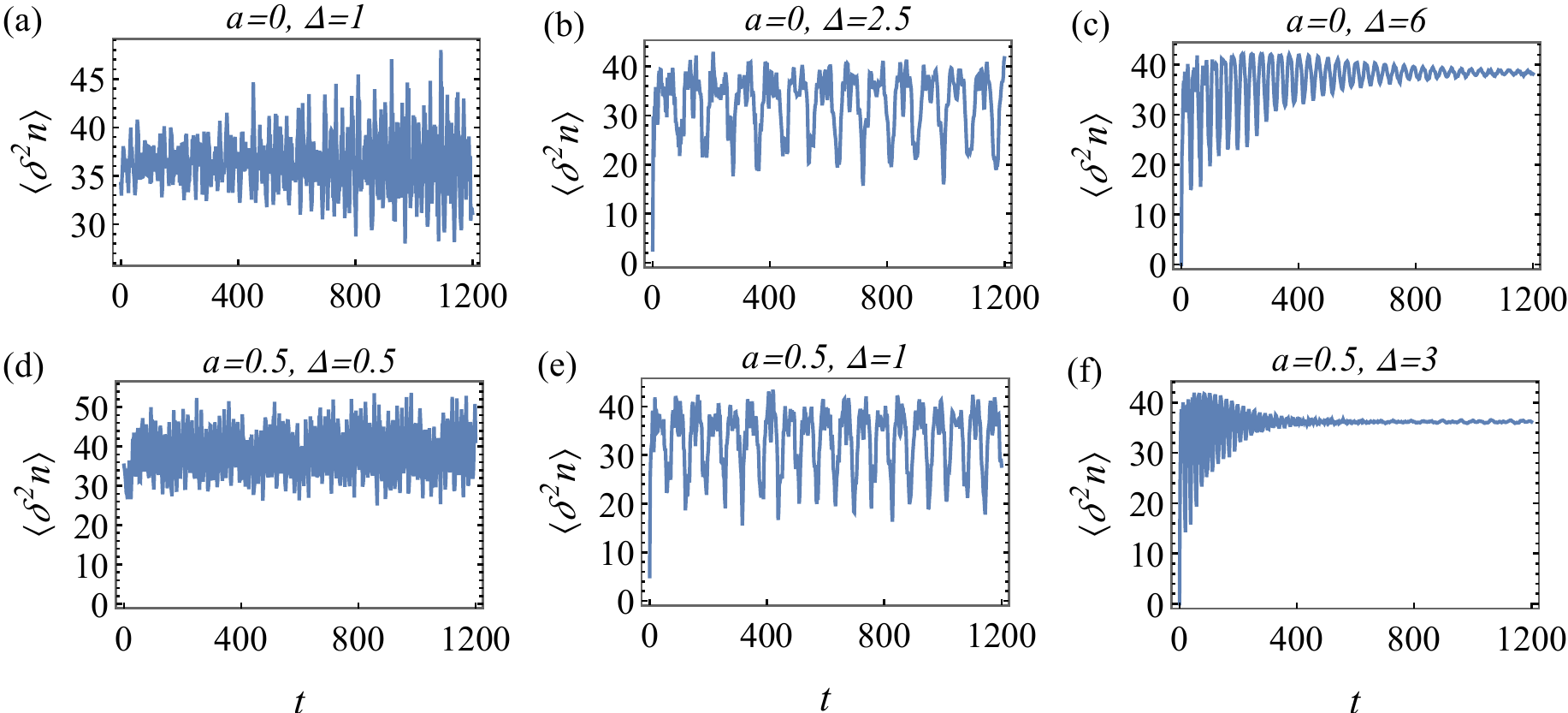}
\caption{(Color online) The  plots of $\langle\delta^2 n\rangle$  for different situations. Except of $a$ and $\Delta$,   $\eta=0.1$, $\omega_c=10$, $s=1$ and $\phi=\pi$ are chosen for these plots. }
\label{fig:variance}
\end{figure*}

To characterize the stability of SP or IPR observed in Fig. \ref{fig:a=0} and \ref{fig:a=0.5}, the variance of the position of excitation within the atomic site,  defined as $\langle\delta^2 n\rangle= \frac{\sum_{n=1}^N \left|\alpha_n(t)\right|^2\left(n-\langle n\rangle\right)^2}{\sum_n \left|\alpha_n(t)\right|^2}$ and $\langle n\rangle=\sum_n\left|\alpha_n(t)\right|^2 n$, is studied for different cases in Fig. \ref{fig:variance}. It is evident that $\langle\delta^2 n\rangle$ can display regular oscillation for $a=0, \Delta=2.5$ and $a=0.5, \Delta=1$. This feature can  also be understood from the coupling induced coherent transition of the two highest  energy levels, as a result of which $\langle\delta^2 n\rangle$ would be determined by the two levels. In contrast, for $a=0, \Delta=1$ and $a=0.5, \Delta=0.5$,  the time evolution of $\langle\delta^2 n\rangle$ becomes irregular. This observation can be attributed to the extensity of the system, by which the excitation would populate uniformly in the atomic sites. However, for $a=0, \Delta=1$ and $a=0.5, \Delta=3$, the system is deeply in the localized phase. Thus, one can find that $\langle\delta^2 n\rangle$ tends to be stable as shown in Fig. \ref{fig:variance}.

\end{document}